\def\modelvtwo{\relax}

\def\deltanu{\relax}

\documentclass{nature}
\usepackage{graphicx}
\usepackage{makecell}
\usepackage{subcaption}
\makeatletter
\let\saved@includegraphics\includegraphics
\AtBeginDocument{\let\includegraphics\saved@includegraphics}

\makeatother

\usepackage[x11names]{xcolor}
\usepackage{amssymb}
\title{X-ray Polarization Reveals the Precessions of the Neutron Star in Hercules X-1}
\begin{document}
%\linenumbers

\author{Jeremy Heyl$^{1}$,
Victor Doroshenko$^{2}$,
Denis Gonz\'alez-Caniulef$^{3}$,
Ilaria Caiazzo$^{4}$,
Juri Poutanen$^{5}$,
Alexander Mushtukov$^{6}$,
Sergey S. Tsygankov$^{5}$,
Demet Kirmizibayrak$^{1}$,
Matteo Bachetti$^{7}$,
George G. Pavlov$^{8}$
% ...
}

\maketitle

\begin{affiliations}
\item University of British Columbia, Vancouver, BC V6T 1Z4, Canada
\item Institut f\"ur Astronomie und Astrophysik, Universität T\"ubingen, Sand 1, 72076 T\"ubingen, Germany
\item Institut de Recherche en Astrophysique et Plan\'etologie, UPS-OMP, CNRS, CNES, 9 avenue du Colonel Roche, BP 44346 31028, Toulouse CEDEX 4, France
\item Division of Physics, Mathematics and Astronomy, California Institute of Technology, Pasadena, CA 91125, USA
\item Department of Physics and Astronomy, 20014 University of Turku, Finland
\item Department of Physics, University of Oxford, Oxford OX1 3RH, UK
\item INAF Osservatorio Astronomico di Cagliari, Via della Scienza 5, 09047 Selargius (CA), Italy
\item Department of Astronomy and Astrophysics, Pennsylvania State University, University Park, PA 16802, USA
\item International Space Science Institute, Hallerstrasse 6, 3012 Bern, Switzerland
\item Instituto de Astrof\'isica de Andaluc\'ia—CSIC, Glorieta de la Astronom\'ia s/n, 18008 Granada, Spain
\item INAF Osservatorio Astronomico di Roma, Via Frascati 33, 00078 Monte Porzio Catone (RM), Italy
\item Space Science Data Center, Agenzia Spaziale Italiana, Via del Politecnico snc, 00133 Roma, Italy
\item Istituto Nazionale di Fisica Nucleare, Sezione di Pisa, Largo B. Pontecorvo 3, 56127 Pisa, Italy
\item Dipartimento di Fisica, Universit\`a di Pisa, Largo B. Pontecorvo 3, 56127 Pisa, Italy
\item NASA Marshall Space Flight Center, Huntsville, AL 35812, USA
\item Dipartimento di Matematica e Fisica, Universit\`a degli Studi Roma Tre, Via della Vasca Navale 84, 00146 Roma, Italy
\item Istituto Nazionale di Fisica Nucleare, Sezione di Torino, Via Pietro Giuria 1, 10125 Torino, Italy
\item Dipartimento di Fisica, Universit\`a degli Studi di Torino, Via Pietro Giuria 1, 10125 Torino, Italy
\item INAF Osservatorio Astrofisico di Arcetri, Largo Enrico Fermi 5, 50125 Firenze, Italy
\item Dipartimento di Fisica e Astronomia, Universit\`a degli Studi di Firenze, Via Sansone 1, 50019 Sesto Fiorentino (FI), Italy
\item Istituto Nazionale di Fisica Nucleare, Sezione di Firenze, Via Sansone 1, 50019 Sesto Fiorentino (FI), Italy
\item INAF Istituto di Astrofisica e Planetologia Spaziali, Via del Fosso del Cavaliere 100, 00133 Roma, Italy
\item ASI - Agenzia Spaziale Italiana, Via del Politecnico snc, 00133 Roma, Italy
\item Science and Technology Institute, Universities Space Research Association, Huntsville, AL 35805, USA
\item Istituto Nazionale di Fisica Nucleare, Sezione di Roma "Tor Vergata", Via della Ricerca Scientifica 1, 00133 Roma, Italy
\item Department of Physics and Kavli Institute for Particle Astrophysics and Cosmology, Stanford University, Stanford, California 94305, USA
\item Astronomical Institute of the Czech Academy of Sciences, Bo\v{c}n\'i II 1401/1, 14100 Praha 4, Czech Republic
\item RIKEN Cluster for Pioneering Research, 2-1 Hirosawa, Wako, Saitama 351-0198, Japan
\item Yamagata University,1-4-12 Kojirakawa-machi, Yamagata-shi 990-8560, Japan
\item Osaka University, 1-1 Yamadaoka, Suita, Osaka 565-0871, Japan
\item International Center for Hadron Astrophysics, Chiba University, Chiba 263-8522, Japan
\item Institute for Astrophysical Research, Boston University, 725 Commonwealth Avenue, Boston, MA 02215, USA
\item Department of Astrophysics, St. Petersburg State University, Universitetsky pr. 28, Petrodvoretz, 198504 St. Petersburg, Russia
\item Department of Physics and Astronomy and Space Science Center, University of New Hampshire, Durham, NH 03824, USA
\item Physics Department and McDonnell Center for the Space Sciences, Washington University in St. Louis, St. Louis, MO 63130, USA
\item Finnish Centre for Astronomy with ESO,  20014 University of Turku, Finland
\item Istituto Nazionale di Fisica Nucleare, Sezione di Napoli, Strada Comunale Cinthia, 80126 Napoli, Italy
\item Universit\'e de Strasbourg, CNRS, Observatoire Astronomique de Strasbourg, UMR 7550, 67000 Strasbourg, France
\item MIT Kavli Institute for Astrophysics and Space Research, Massachusetts Institute of Technology, 77 Massachusetts Avenue, Cambridge, MA 02139, USA
\item Graduate School of Science, Division of Particle and Astrophysical Science, Nagoya University, Furo-cho, Chikusa-ku, Nagoya, Aichi 464-8602, Japan
\item Hiroshima Astrophysical Science Center, Hiroshima University, 1-3-1 Kagamiyama, Higashi-Hiroshima, Hiroshima 739-8526, Japan
\item University of Maryland, Baltimore County, Baltimore, MD 21250, USA
\item NASA Goddard Space Flight Center, Greenbelt, MD 20771, USA
\item Center for Research and Exploration in Space Science and Technology, NASA/GSFC, Greenbelt, MD 20771, USA
\item Department of Physics, The University of Hong Kong, Pokfulam, Hong Kong
\item Universit\'e Grenoble Alpes, CNRS, IPAG, 38000 Grenoble, France
\item Harvard-Smithsonian Center for Astrophysics, 60 Garden St, Cambridge, MA 02138, USA
\item INAF Osservatorio Astronomico di Brera, Via E. Bianchi 46, 23807 Merate (LC), Italy
\item Dipartimento di Fisica e Astronomia, Universit\`a degli Studi di Padova, Via Marzolo 8, 35131 Padova, Italy
\item Dipartimento di Fisica, Universit\`a degli Studi di Roma "Tor Vergata", Via della Ricerca Scientifica 1, 00133 Roma, Italy
\item Department of Astronomy, University of Maryland, College Park, Maryland 20742, USA
\item Mullard Space Science Laboratory, University College London, Holmbury St Mary, Dorking, Surrey RH5 6NT, UK
\item Anton Pannekoek Institute for Astronomy \& GRAPPA, University of Amsterdam, Science Park 904, 1098 XH Amsterdam, The Netherlands
\item Guangxi Key Laboratory for Relativistic Astrophysics, School of Physical Science and Technology, Guangxi University, Nanning 530004, China
\end{affiliations}

\begin{abstract}
%Hercules X-1 is in many ways the prototypical X-ray pulsar.  As the second to be discovered, its emission varies on three distinct timescales\cite{1972ApJ...174L.143T}:  it rotates every 1.2~seconds, it is eclipsed by its companion each 1.7~days, and it exhibits a superorbital period of 35~days. Over this 35-day-period, the accretion disc precesses and periodically blocks the emission regions near the neutron star\cite{2000ApJ...539..392S}. We present compelling evidence from X-ray polarization measurements with the Imaging X-ray Polarimetry Explorer\cite{2022JATIS...8b6002W} (IXPE) that this 35-day-period is set by the free precession of the neutron star itself, which has three distinct eigenvalues of the moment of inertia\cite{2022MNRAS.513.3359K} that differ fractionally by a few parts per ten million.  Furthermore, we find evidence that longer-acting torques are changing the spin geometry of the neutron star.
In an accreting X-ray pulsar, a neutron star accretes matter from a stellar companion through an accretion disk.  The high magnetic field of the rotating neutron star disrupts the inner edge of the disc, funneling the gas to flow onto the magnetic poles on its surface. Hercules X-1 is in many ways the prototypical X-ray pulsar; it shows persistent X-ray emission and it resides with its companion HZ Her, a two-solar-mass star, at about 7~kpc from Earth\cite{2021AJ....161..147B}. Its emission varies on three distinct timescales\cite{1972ApJ...174L.143T}: the neutron star rotates every 1.2~seconds, it is eclipsed by its companion each 1.7~days, and the system exhibits a superorbital period of 35~days which has remained remarkably stable since its discovery\cite{1973ApJ...184..227G}. Several lines of evidence point to the source of this variation as the precession of the accretion disc\cite{1991ApJ...378..696P,2000ApJ...539..392S}, the precession of the neutron star\cite{1972Natur.239..325B,1986ApJ...300L..63T} or both\cite{2009A&A...494.1025S}. Despite the many hints over the past fifty years, the precession of the neutron star itself has yet not been confirmed or refuted. We here present X-ray polarization measurements with the Imaging X-ray Polarimetry Explorer\cite{2022JATIS...8b6002W} (IXPE) which probe the spin geometry of the neutron star.  These observations provide direct evidence that the 35-day-period is set by the free precession of the neutron star crust, which has the important implication that its crust is somewhat asymmetric fractionally by a few parts per ten million\cite{2022MNRAS.513.3359K}. Furthermore, we find indications that the basic spin geometry of the neutron star is altered by torques on timescale of a few hundred days. 

\end{abstract}

\begin{figure*}[ht]
\centering
  \includegraphics[width=\textwidth]{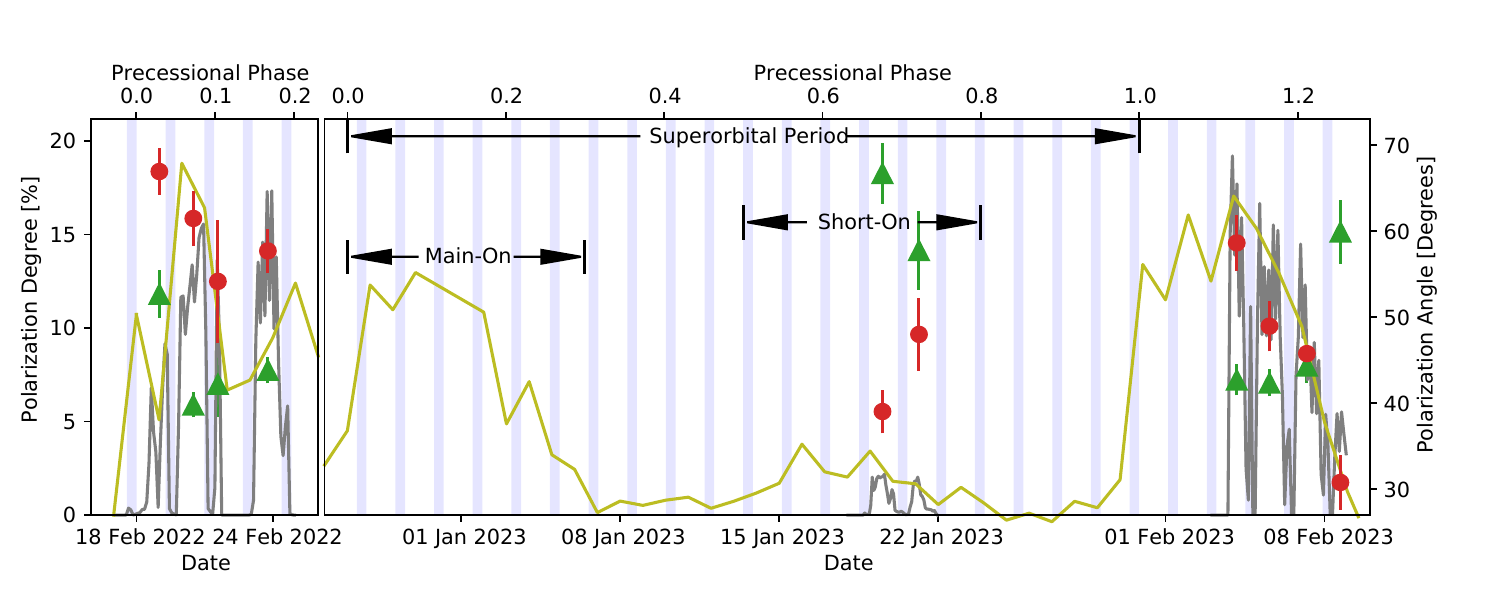}
  \caption{
  \textbf{Evolution of the observed flux, PA and PD from Her X-1.} Flux depicted by grey curve from IXPE and olive from Swift, PD by green triangles using the left $y-$axis) and PA by red circles using the right $y-$axis, time and phase
of the 35-day super-orbital precession cycle on the two $x-$axes. The turn-on time 2022 February 18 is estimated from the IXPE data, and those of 2022 December 26 and 2023 January 30 are estimated from the Swift data. The main-on lasts from precessional phase 0 to 0.3 (2022 December 26 to 8 January 2023), and the short-on lasts from 0.5 to 0.8 (15-25 January 2023). These phases repeat every 34.84 days (the superorbital period).
The reported values and the uncertainties correspond to the mean values and $1\sigma$ (68\%) confidence
intervals and agree with the previously reported results\cite{2022NatAs...6.1433D,2023ApJ...948L..10G}. The shaded regions depict the intervals when the companion star and its accretion stream eclipse the neutron star.  We define the superorbital phase of zero to coincide with the beginning of the main-on\cite{2022NatAs...6.1433D,2000ApJ...539..392S}.}\label{fig:flux}
\end{figure*}

Hercules X-1 was observed by IXPE\cite{2022JATIS...8b6002W} on 2022 February 17-24 (255~ks), at the beginning of the 35 d precession cycle, the so-called ``main-on'' state\cite{2022NatAs...6.1433D} and again in 2023 January and February during the ``short-on'' (148~ks) and the ``main-on'' state (245~ks) as shown in Fig.~\ref{fig:flux}. The gas-pixel detector on IXPE registers the arrival time, sky position, and energy for each X-ray photon and uses the photoelectric effect to provide an estimate of the position angle of each photon\cite{2021AJ....162..208S}. During each observation, photon arrival events registered between energies of two and eight keV, within 52 arcseconds of the position of source were extracted for analysis. 
% A second annular region centered on the source of inner radius 78 arcseconds and outer radius of 130 arcseconds formed the background region.  For each observation the number of photons in the background region was less than 1.6\% of those in the source region, so we ignore the background for our further polarimetric analysis.  
We do not subtract background for the analysis as a bright source such as Hercules X-1 dominates the background over a very large region of the detector\cite{Di_Marco2023}.
The times of the photon arrivals were corrected for the motion of IXPE around the barycenter of the Solar System, and the orbit of the neutron star about its companion to be in a frame of reference moving with the neutron star. The details of the data reduction are outlined in the methods section of Doroshenko et al.\cite{2022NatAs...6.1433D}.

\begin{figure*}[ht]
\centering
%2022 February\hfill2023 January\hfill2023 February
  \includegraphics[width=0.9\textwidth]{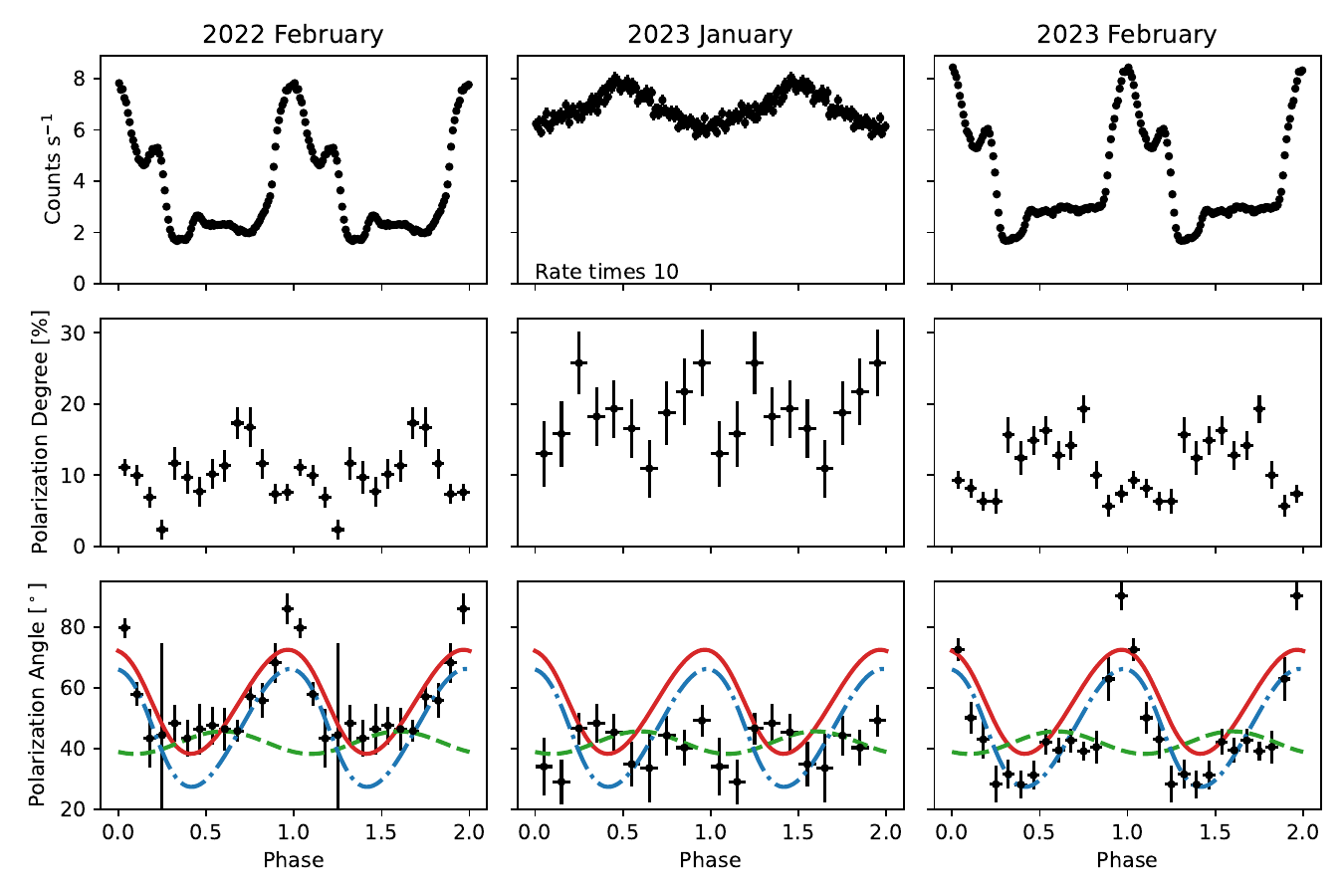}
    \caption{\textbf{IXPE Observations of Hercules X-1 as a Function of Spin Phase.} 
    The top, middle and bottom rows show with black errorbars: the IXPE count rate, the polarization degree, and the polarization angle, respectively.
    Left, middle and right columns are for the first main-on (2022 February), the short-on (2023 January), and the second main-on (2023 February), respectively.  Error bars indicate $\Delta \log L=1/2$ and are equivalent to one-sigma. The solid red, dashed green and dot-dashed blue curves in the bottom row depict the best-fitting RVM models for the first main-on, the short-on and the second main-on respectively.} 
    \label{fig:obs}
\end{figure*}

The analysis focuses on the geometry of the spinning neutron star over the pulsar rotation period and the superorbital period. We use a maximum likelihood technique\cite{gonzalezcaniulef_unbinned} to determine the mean polarization properties averaged over the pulsar orbit as shown in Fig.~\ref{fig:flux}, and as a function of pulse phase during the different superorbital states of the system in Fig.~\ref{fig:obs}.  We track the observed polarization angle as the star rotates to determine the geometry of the spinning neutron star and study the geometry as the system evolves through the superorbital period.  We fix the phase of zero in the two main-on epochs to the peak of the X-ray light curve.  For the short-on, we fix the phase of one half to the peak of the light curve 
in agreement with previous work\cite{2000ApJ...539..392S}. Our results do not depend on the choice of phasing among the epochs. As already apparent in Fig.~\ref{fig:flux}, the source appears much fainter during the short-on than during the main-on (this is often attributed to occultation by the accretion disk\cite{2000ApJ...539..392S}), so the count rate for the short-on in Fig.~\ref{fig:obs} must be inflated by a factor of ten to be visible clearly.  The polarization degree in the two main-on epochs is about five to fifteen percent and in the short-on it is about two times larger.  The pulse profile and polarization degree are consistent between the two main-on epochs\cite{1990ApJ...348..641S}.  More dramatic changes are apparent when one examines the polarization angle as a function of spin phase.  The evolution of the polarization angle over the spin period during the two main-on epochs are nearly identical up to an offset. Over the course of the year, between the two main-on epochs, the mean value of the polarization angle has decreased by $9^\circ\!\!.2 \pm 1^\circ\!\!.8$.  The variation of the polarization angle during the short-on, on the other hand, is much smaller than during the main-on.

 %                       _           _     
 %     /\               | |         (_)    
 %    /  \   _ __   __ _| |_   _ ___ _ ___ 
 %   / /\ \ | '_ \ / _` | | | | / __| / __|
 %  / ____ \| | | | (_| | | |_| \__ \ \__ \
 % /_/    \_\_| |_|\__,_|_|\__, |___/_|___/
 %                          __/ |          
 %                         |___/           

Although the magnetic field structure in the emission region of the pulsar is expected to be complicated\cite{2013MNRAS.435.1147P},
the polarization angles of photons coming from different parts of the emission region are expected to follow the magnetic field direction as they propagate in the highly magnetized plasma surrounding the X-ray pulsar. Even at large distances from the neutron star, where the plasma does not affect the radiation, vacuum birefringence causes the polarized radiation in the magnetosphere to propagate in the ordinary (O) and extraordinary (X) modes which represent oscillations of the electric field parallel and perpendicular to the plane formed by the local magnetic field and the photon momentum\cite{1978SvAL....4..117G,1979JETP...49..741P}, and propagation in the normal modes continues within the so-called polarization limiting radius\cite{2000MNRAS.311..555H}. For typical X-ray pulsars, this radius is
estimated to be about thirty stellar radii\cite{2018Galax...6...76H}, where the field is dominated by the dipole component, so the polarization measured at the telescope is expected to be either parallel or perpendicular to the instantaneous projection of the magnetic dipole axis of the star onto the plane of the sky. For this reason, the modulation of the polarization angle (PA) with phase is decoupled from the evolution in polarization degreee (PD) and intensity and should follow the rotating vector model (RVM)\cite{1969ApL.....3..225R,2020A&A...641A.166P,gonzalezcaniulef_unbinned} 
\begin{equation}
    \tan(\textrm{PA}-\chi_{\rm p}) = \frac{\sin\theta\sin(\phi-\phi_0)}{\cos i_{\rm p} \sin\theta\cos(\phi-\phi_0)-\sin i_{\rm p} \cos\theta } ,
\end{equation}
where $i_{\rm p}$ is the inclination of the angular velocity with respect to the line of sight, $\chi_{\rm p}$ the position angle of the rotation axis in the plane of the sky, $\theta$ is the inclination of the magnetic dipole to the spin axis, $\phi$ is the spin phase, and $\phi_0$ is the phase of the rotation when the magnetic dipole axis is closest to the line of sight. 

Neutron stars probably do not rotate as rigid bodies\cite{1987A&A...185..196A}.  The magnetic field, whose geometry we measure through the polarization, is anchored to the rigid crust of the neutron star and passes through the core of the neutron star where it interacts with the core superfluid; consequently, the measured polarization angle through the RVM probes the instantaneous rotation axis of the neutron-star crust through the parameters $\chi_{\rm p}$ and $i_{\rm p}$, and the location of the magnetic pole relative to the instantaneous rotation axis through the parameters $\theta$ and $\phi_0$.  If the neutron-star crust is prolate, oblate, or triaxial and the spin axis does not coincide with a symmetry axis of the crust, the neutron star can precess even in the absence of torques, so the values of $\chi_{\rm p}$, $i_{\rm p}$, $\theta$ and $\phi_0$ can change with time as well, over the precession period of the neutron star.  This would change the evolution of the polarization angle with spin phase.  We can infer that the neutron star is nearly symmetric to within a few parts per ten million from the ratio of the spin period of 1.2~seconds to the potential precession period of 35~days (the superorbital period)\cite{Landau1976Mechanics}; if the angular momentum of the crust is conserved, the angles $\chi_{\rm p}$ and $i_{\rm p}$ would remain constant also to within a few parts per ten million, and therefore any observed changes in these angles are a hallmark of a net torque on the neutron-star crust.  The values of $\theta$ and $\phi_0$, on the other hand, can change measurably even in the absence of torques as the spin axis moves across the surface of the neutron star.  

\begin{figure*}[ht]
  \centering
  \ifx\origschematic\undefined
    \includegraphics[width=0.9\textwidth]{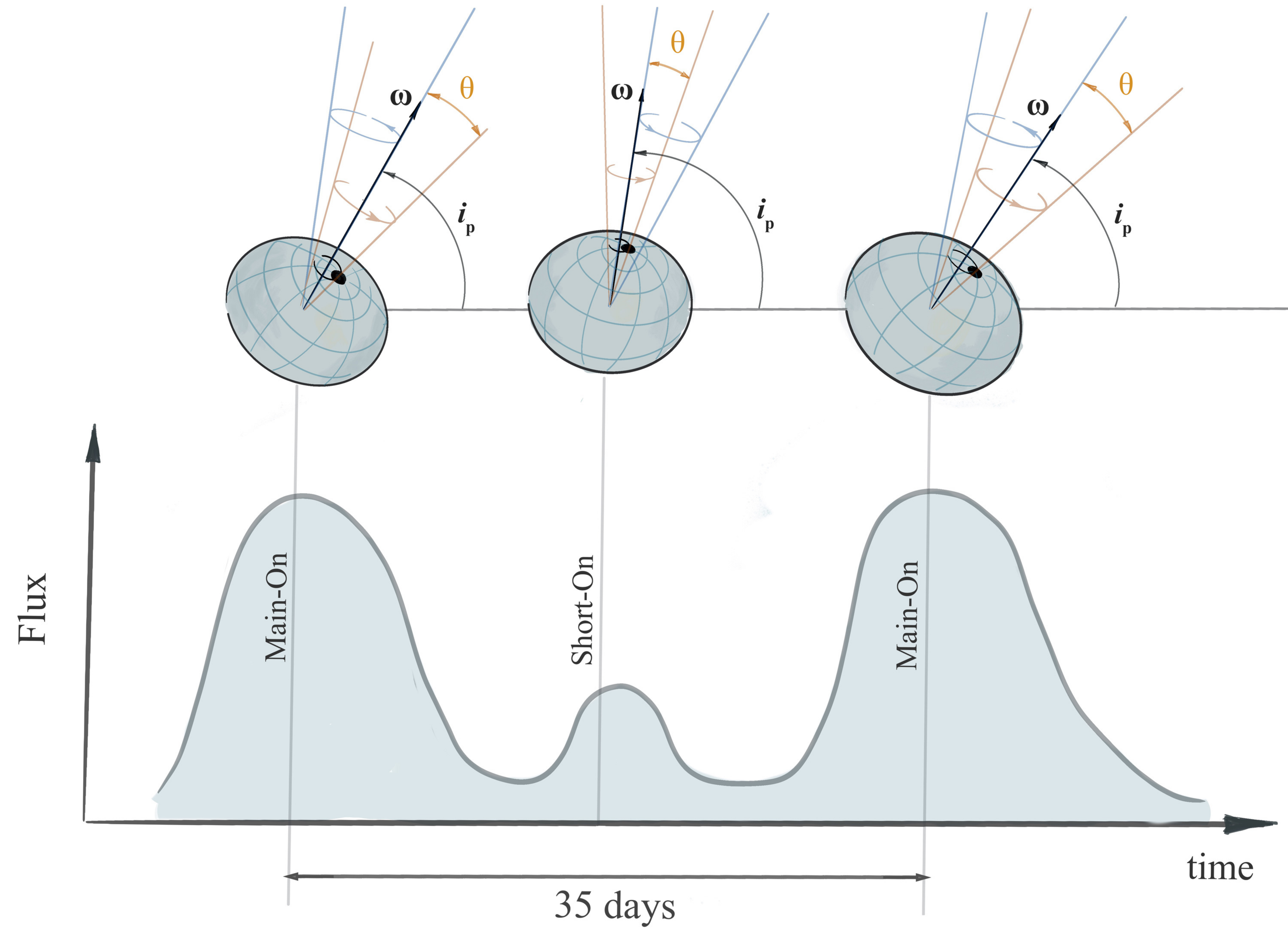}
  \else
    \includegraphics[width=0.9\textwidth]{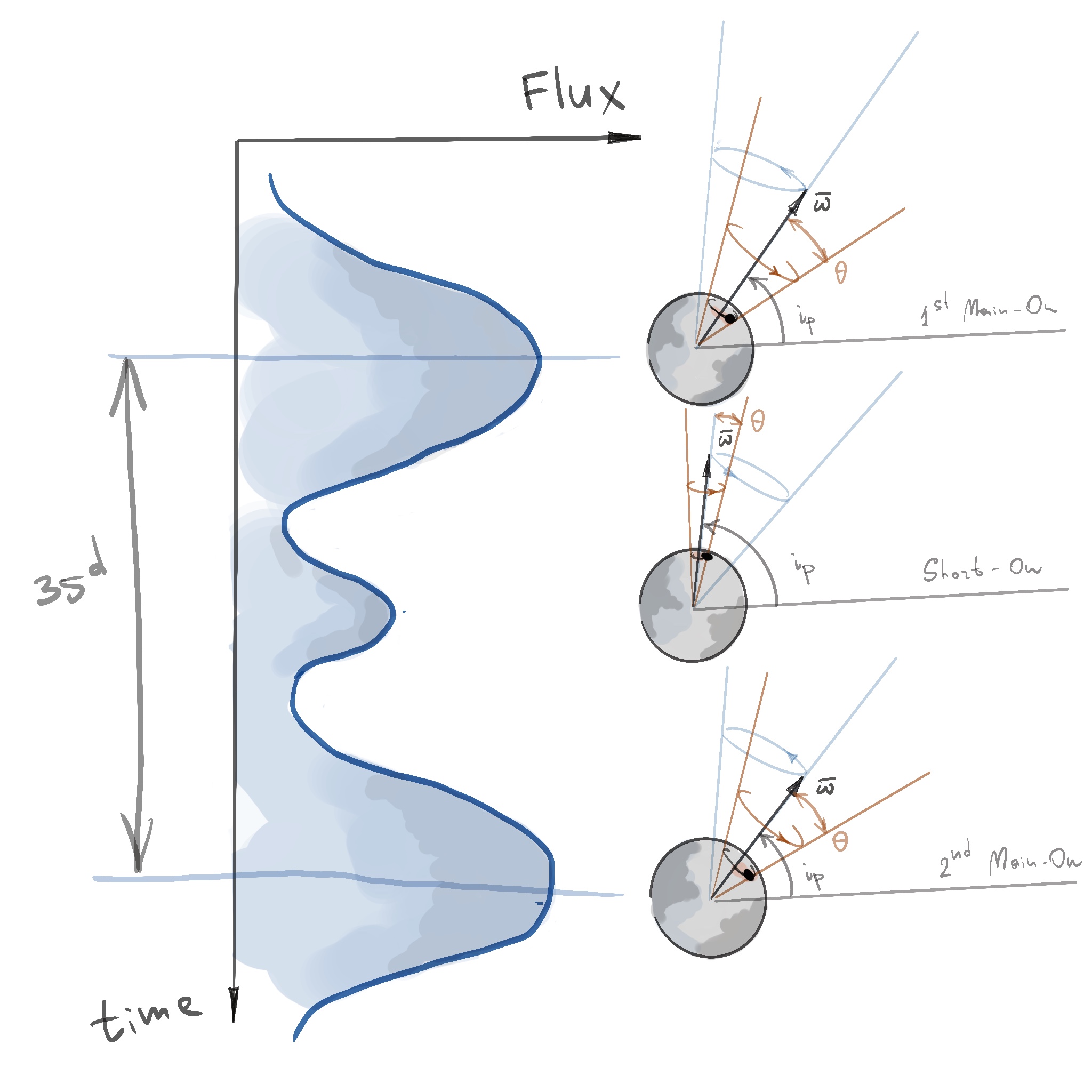}
   \fi
    \caption{\textbf{Schematic showing the varying geometry through the precession of Hercules X-1.} The dark spot on the surface of the neutron star indicates the location of the magnetic pole.}
 \label{fig:schematic}
\end{figure*}

To quantify the observed changes, we fit the photon arrival times and photo-electron angles directly to an RVM of the polarization angle\cite{gonzalezcaniulef_unbinned}. The colored curves in Fig.~\ref{fig:obs} depict the best-fitting RVM models to the measured photo-electron trajectories.  The best-fitting parameters for these fits are given in Table~\ref{tab:rvm_param}, and the covariances are shown in Extended Data Figures~\ref{fig_extended:epoch1}~through~\ref{fig_extended:epoch3}.  Fig.~\ref{fig:schematic} depicts the varying geometry schematically. The values for the first main-on epoch agree with those measured by Doroshenko et al.\cite{2022NatAs...6.1433D} within the uncertainties.  To examine whether the parameters change over the course of a single main-on phase, we fit only the data during the first two orbital periods (Early Epoch 1) and the final orbit (Late Epoch 1), and present these results as well.  During the first main-on, the RVM parameters do not vary substantially.  However, the differences are significant among the three epochs: first main-on, short-on, and second main-on.  In particular, as indicated by the phase-resolved polarimetry itself, the position angle of the spin axis on the sky ($\chi_{\rm p}$) has changed by nine degrees ($8^\circ\!\!.6\pm2^\circ\!\!.2$) between the two main-on epochs while the other parameters remain the same within the uncertainties.  This indicates the angular momentum of the crust has changed from 2022 February to 2023 February.

Furthermore, between the short-on and the latter main-on, there is strong evidence ($p\approx0.008$) that the value of the angle between the spin axis and the magnetic dipole ($\theta$) is larger during the main-on than during the short-on: $16^\circ\!\!.0_{-4^\circ\!\!.3}^{+3^\circ\!\!.1}$ versus $3^\circ\!\!.7_{-1^\circ\!\!.9}^{+2^\circ\!\!.6}$.  The larger value of $\theta$ is reflected in the larger amplitude of modulation in the polarization angle that we noted earlier.  In principle, a larger contribution of scattered X-rays could reduce the swing in polarization during the short-on as well, but this would also reduce the polarization degree below what is observed; furthermore,  observations of the winds and corona through the superorbital period\cite{2021A&A...648A..39S,Kosec23} support the conclusion that scattering does not contribute much to the observed radiation during this epoch. The change in the value of $\theta$ between the short-on and the second main-on indicates that the spin axis of the crust has moved relative to the magnetic axis of the neutron star; as we do not see evidence of a change of $i_M$ or $\chi_M$ over this short eighteen-day period, we argue that a net torque is not required to account for this precession; it is approximately free.  Furthermore, the torque that would be required to cause the spin axis to move through a symmetric crust by twelve degrees over eighteen days exceeds that supplied by the disk by a factor of three\cite{1980SvAL....6...14L,1999ApJ...524.1030L} and would very likely also cause the orientation of the spin axis to change with the respect to the sky by a comparable amount which is not observed. The approximately free precession of modestly asymmetric crust over the 35-day superorbital period remains as the most conservative explanation for the observed short-term changes in the geometry.

 %  __  __           _      _     
 % |  \/  |         | |    | |    
 % | \  / | ___   __| | ___| |___ 
 % | |\/| |/ _ \ / _` |/ _ \ / __|
 % | |  | | (_) | (_| |  __/ \__ \
 % |_|  |_|\___/ \__,_|\___|_|___/

% Suggestion for the next paragraph: The observed evolution in polarization angle shows that the neutron star is precessing over the 35-day superorbital period; however, such evolution can be reproduced by assuming different geometries and shapes of the neutron star itself. We here analyze two possible cases for the neutron star's shape: a prolate model and a triaxial model. Postnov et al.\cite{2013MNRAS.435.1147P} had proposed a prolate model to account for the variation in pulse profiles, and Kolesnikov, Shakura and Postnov\cite{2022MNRAS.513.3359K} used a triaxial model to explain the variation in the spin frequency of the star\cite{2022MNRAS.513.3359K}. The key observables for both cases are the minimum and maximum angles between the instantaneous spin axis and the magnetic pole ($\theta$) which determine broadly how the polarization will evolve through the processional cycle. Also, we assume that the angular momentum of the star makes an angle $i_M=110^\circ$ with respect to the line of sight as this is the approximate inclination of the orbit\cite{2022NatAs...6.1433D,2014ApJ...793...79L}.  We varied this angle from $60^\circ$ to $120^\circ$ and verified that changes in this angle only make minor changes to the predicted polarization angles.
Several models have been proposed for the free precession of the neutron star in Hercules X-1 to explain the superorbital period.  Postnov et al.\cite{2013MNRAS.435.1147P} propose a prolate model to account for the variation in pulse profiles, and Kolesnikov, Shakura and Postnov\cite{2022MNRAS.513.3359K} use a triaxial model to explain the variation in the spin frequency of the star\cite{2022MNRAS.513.3359K}. We first focus on these specific models. The key observables are the minimum and maximum angles between the instantaneous spin axis and the magnetic pole ($\theta$) which determine broadly how the polarization will evolve through the precessional cycle. In all cases, we assume that the angular momentum of the star makes an angle $i_M=110^\circ$ with respect to the line of sight as this is the approximate inclination of the orbit\cite{2022NatAs...6.1433D,2014ApJ...793...79L}.  We varied this angle from $60^\circ$ to $120^\circ$ and verified that changes in this angle only make minor changes to the predicted polarization angles. 

In the prolate model, the ratio of the spin period to the superorbital period is simply the relative difference between the moments of inertia, and the two additional parameters are the time during the precession when the spin axis passes closest to the magnetic axis, and the angle between the spin and the symmetry axis of the star (the spin misalignment angle).  Two prolate models are depicted in the upper two panels of Fig.~\ref{fig:models}. The upper-left one has a spin misalignment of $\alpha=50^\circ$ and an angle of $\theta_B=30^\circ$ between the symmetry axis and the magnetic pole (magnetic obliquity), as proposed by Postnov et al.\cite{2013MNRAS.435.1147P}.  These angles, along with the precession and rotation angles ($\psi$ and $\varphi$), are depicted in Extended Data Figure~\ref{fig_extended:Euler}.  This model, in blue in the figure, predicts a variation in the polarization angle throughout the precession that is larger than observed (superimposed colored 
\ifx\modelvtwo\undefined
curves).
\else
regions).
\fi
To achieve better agreement with the observed polarization angles, the spin misalignment and magnetic obliquity must be reduced to $11^\circ$ and $8^\circ$ respectively (upper-right panel). These angles depend modestly on the assumed value of $i_M$: both are smaller if $i_M<50^\circ$ or $i_M>130^\circ$.

Although the triaxial models cannot be solved using elementary functions, the evolution can be expressed in closed form\cite{Landau1976Mechanics,2022MNRAS.513.3359K,1998A&A...331L..37S} (see the Methods section).  In particular, the precession period is determined by the relative difference between the largest and smallest eigenvalues of the moment of inertia\cite{2022MNRAS.513.3359K}, taken to be $6.7\times 10^{-7}$.  The relative difference between the second largest and smallest eigenvalues is taken\cite{2022MNRAS.513.3359K} to be $2.7\times 10^{-7}$.  We examine two pairings for the minimum spin misalignment angle and the magnetic obliquity: $50^\circ$ and $30^\circ$, as postulated by Kolesnikov, Shakura and Postnov\cite{2022MNRAS.513.3359K}, in the bottom-left panel, and $9.5^\circ$ and $7.5^\circ$ in the bottom-right panel.  In both cases, the magnetic axis lies toward the intermediate axis of the neutron star.  In the bottom-left panel, we see that the variation in polarization angle is larger throughout the entire precession period than observed during either the short-on or the main-on. As with the prolate model, better agreement is achieved by reducing the misalignment angle and the magnetic obliquity which results in the models in the bottom-right panel which can account for the observed evolution in the polarization over the superorbital period.  Furthermore, frequency variations reported by Kolesnikov, Shakura and Postnov\cite{2022MNRAS.513.3359K} are consistent with free precession models, and the models that we present here are also consistent with those variations
\ifx\deltanu\undefined
.
\else
as depicted in Extended Data Fig.\ref{fig:deltanu}.
\fi

%Furthermore, these smaller values of magnetic obliquity and misalignment in the triaxial model can reproduce the changes in spin frequency observed for the pulsar\cite{2022MNRAS.513.3359K} as depicted in Fig.\ref{fig:deltanu}.
%  _____  _                        _             
 % |  __ \(_)                      (_)            
 % | |  | |_ ___  ___ _   _ ___ ___ _  ___  _ __  
 % | |  | | / __|/ __| | | / __/ __| |/ _ \| '_ \ 
 % | |__| | \__ \ (__| |_| \__ \__ \ | (_) | | | |
 % |_____/|_|___/\___|\__,_|___/___/_|\___/|_| |_|

 %  _____  _     _    
 % |  __ \(_)   | |   
 % | |  | |_ ___| | __
 % | |  | | / __| |/ /
 % | |__| | \__ \   < 
 % |_____/|_|___/_|\_\

The neutron star crust and the core are strongly coupled and, in the absence of torques, relative motion between them dissipates on a timescale of minutes to hours\cite{1987A&A...185..196A}.  The coupling between the crust and the superfluid within the crust is mediated through the pinning and unpinning of vortices on nuclei in the crust in a process called vortex creep. Typically, the lag between the crust and the crust superfluid is larger than between the crust and the core superfluid.  These dissipative processes will damp the wobbling of the crust over a timescale of a few hundred years in the absence of external torques. On the other hand, the external torques, supplied by accretion in the case of Hercules X-1, can support an equilibrium where the external torques on the crust balance both the torques between the crust and the core superfluid and the larger torque between the crust and the crust superfluid, so the crust itself can exhibit a precession approximately free of torques.  In fact an external torque, as the one exerted by accretion, can drive the crust to precess and the amplitude of the precession depends on the long-term behaviour of the external torque. This torque increases with the sine of the misalignment angle between the spin axis and the figure axis of the neutron star crust, so larger misalignments require larger torques to maintain, and a stable equilibrium with approximately free precession of the crust can be achieved. For Hercules X-1, the total precessional torque exerted magnetically by the accretion disk on the neutron-star crust is\cite{1980SvAL....6...14L,1999ApJ...524.1030L}
$$
N^{{\textrm{\scriptsize prec}}}_{\textrm{\scriptsize tot}} \approx 2 \sin^2 \theta \sin2\beta \left(\frac{B_p}{3\times 10^{12} \textrm{G}}\right)^2 \left ( \frac{R_m}{1000\textrm{km}} \right)^{-3} \times 10^{36}~\textrm{dyne cm}
$$
where $B_p$ is the magnetic field strength at the pole of the neutron star, $R_m$ is radius of the inner edge of the accretion disk and $\beta$ is the misalignment angle between the angular momentum of the spin of the neutron star and the angular momentum of the orbit. Furthermore, the magnetic field exerts an additional torque on the disk that tends to warp it\cite{1999ApJ...524.1030L}
$$
N^{{\textrm{\scriptsize warp}}}_{\textrm{\scriptsize tot}} \approx 0.2 \cos^2 \theta \sin2\beta \left(\frac{B_p}{3\times 10^{12} \textrm{G}}\right)^2 \left ( \frac{R_m}{1000\textrm{km}} \right)^{-3} \times 10^{36}~\textrm{dyne cm}.
$$
Both of these torques vanish if the accretion disk lies above the rotational equator of the neutron star ($\beta=0$); however, the misalignment between the neutron-star spin and the orbit is measured to be at least 24 degrees\cite{2022NatAs...6.1433D} so $\sin2\beta>0.7$, so both the precessional and warping torques are finite and vary with the angle between the spin axis and the magnetic axis ($\theta$).

As the crust executes its precessional motion, the torque exerted on the disk by the magnetic field of the neutron star will also vary on the same timescale because the angle between the spin axis and the magnetic axis ($\theta$) changes. These torques excite the precession and warping of the disk that are invoked to explain the evolution of the system through the superorbital phase, and their variation through the changing value of $\theta$ over the precession of the neutron-star crust sets the superorbital timescale. From the point of view of the neutron star, these torques serve to maintain the approximate free precession of the crust against internal dissipation which would damp the precession within hundreds of years\cite{1987A&A...185..196A}.

 %
 % |__   __|                             
 %    | | ___  _ __ __ _ _   _  ___  ___ 
 %    | |/ _ \| '__/ _` | | | |/ _ \/ __|
 %    | | (_) | | | (_| | |_| |  __/\__ \
 %    |_|\___/|_|  \__, |\__,_|\___||___/
 %                    | |                
 %                    |_|    
 %
Between the first and second main-on observed by IXPE, the position angle of the spin axis on the sky has shifted by about nine degrees.  If we take the moment of inertia of the neutron-star crust to be $10^{43}$ g cm$^{2}$ (about one hundredth of the entire neutron star)\cite{2015PhRvC..91a5804S}, the mean torque over the 355~days between the observations is $2.6\times 10^{35}$~dyne cm, a factor of ten smaller than the precessional torque supplied by the disk with an inner radius of 1000~km. Although in equilibrium the net torque on the crust vanishes, the evolution of the spin of Hercules X-1 is complicated.  The amplitude of the precession and therefore the magnitude of the internal torques necessary to maintain the precession depend on the mean value of the accretion torques over the damping time of several hundred years; whereas the external torques depend on the current accretion; if these two differ by about ten percent, the resulting net torque on the crust could result in the observed net precession of the spin axis of the crust on the sky.  In fact continued observations of Hercules X-1 could constrain the mean accretion rate over the past few hundred years as well as details of the crust-superfluid coupling.

 %  ______ _             _  __          __           _     
 % |  ____(_)           | | \ \        / /          | |    
 % | |__   _ _ __   __ _| |  \ \  /\  / /__  _ __ __| |___ 
 % |  __| | | '_ \ / _` | |   \ \/  \/ / _ \| '__/ _` / __|
 % | |    | | | | | (_| | |    \  /\  / (_) | | | (_| \__ \
 % |_|    |_|_| |_|\__,_|_|     \/  \/ \___/|_|  \__,_|___/
                                                         
Polarization measurements with IXPE reveal that the polarization angle as a function of spin phase changes over the superorbital period and also over longer timescales.  We interpret these observations as signatures of the free precession and forced precession of the crust of the neutron star in Hercules~X-1, confirming that the neutron star itself provides the clock for the superorbital period\cite{2009A&A...500..891S} and revealing that torques also change the angular momentum of the crust on a year-long timescale. This forced-precession may account for the anomalous lows in Hercules X-1 and modulation of the superorbital period that occur on a five-year timescale\cite{2009A&A...494.1025S}. Further polarization observations of Hercules~X-1 will probe the interior of the neutron star, in particular the coupling between the crust and the superfluid, as well as the accretion dynamics, and polarization observations of other X-ray binaries with superorbital periods, a common phenomenon in these sources, may verify their underlying clocks.

 %  _______    _     _           
 % |__   __|  | |   | |          
 %    | | __ _| |__ | | ___  ___ 
 %    | |/ _` | '_ \| |/ _ \/ __|
 %    | | (_| | |_) | |  __/\__ \
 %    |_|\__,_|_.__/|_|\___||___/

\clearpage

\begin{table}
\centering
\caption{\textbf{Best-fitting RVM Parameters}. The observations for the first main-on were from 2022 February 17 to 24, and the early portion corresponds to the first four days of this segment.  The short-on was observed from 2023 January 18 to 21, and the second main-on was from 2023 February 3 to 8. The second-to-last column gives the median precession phase for the observation.   The final column indicates the quality of the fit.  When the data is consistent with the model, the log-likelihood ($\log L$) is normally distributed, and the fit quality in the last column is given in terms the best-fitting log-likelihood compared the expected value where positive values indicate better than expected values.} \label{tab:rvm_param}
\small{
\begin{tabular}{l|ccccccc}
\hline
\hline
& 
Mean PD &  $\chi_{\rm p}$ &  $\theta$ & $i_{\rm p}$ & $\phi_0/2\pi$ & Prec.\ Phase & $\Delta\log L$ \\
% \multicolumn{1}{c}{Mean PD} &
% \multicolumn{1}{c}{$i_{\rm p}$} &
% \multicolumn{1}{c}{$\theta$} &
% \multicolumn{1}{c}{$\chi_{\rm p}$} & 
% \multicolumn{1}{c}{$\phi_0$} &
% \multicolumn{1}{c}{Prec.\ Phase} 
%\\
&  [\%] & [deg] &   [deg] &  [deg] &  &  & [$\sigma$] \\ 
% \multicolumn{1}{c}{(\%)} &
% \multicolumn{1}{c}{(deg)} & 
% \multicolumn{1}{c}{(deg)} &
% \multicolumn{1}{c}{(deg)} & 
% \multicolumn{1}{c}{(\%)} &
% \multicolumn{1}{c}{(\%)} 
%\\
\hline  
\hline  
First Main-On  & $ 9.5 \pm 0.5$ & $55.4\pm 1.6$ & $14.5^{+3.0}_{-4.0}$ & $58^{+28}_{-22}$ & $0.19^{+0.03}_{-0.02}$ &  $0.088$ & $-1.52$ \\
~~Early        & $ 8.6 \pm 0.6$ & $57.9\pm 2.1$ & $16.3^{+3.5}_{-4.1}$ & $64^{+25}_{-22}$ & $0.19^{+0.03}_{-0.02}$ &  $0.073$ & $+0.07$ \\
~~Late         & $ 9.3 \pm 0.7$ & $52.2\pm 2.7$ & $15.9^{+3.6}_{-4.0}$ & $85^{+35}_{-37}$ & $0.22^{+0.05}_{-0.05}$ & $0.162$ &  $+0.05$\\
Short-On       & $17.8 \pm 1.4$ & $41.9\pm 2.2$ & $ 3.7^{+2.6}_{-1.9}$ & $90^{+30}_{-30}$ & $0.85^{+0.18}_{-0.20}  $ & $0.687$ & $-0.19$\\
Second Main-On & $ 9.1 \pm 0.5$ & $46.8\pm 1.5$ & $16.0^{+3.1}_{-4.3}$ & $56^{+24}_{-20}$ & $0.20^{+0.02}_{-0.02}$ & $0.159$ & $+0.48$ \\
\hline
\end{tabular}}
\end{table}

 %
 %  ______ _                           
 % |  ____(_)                          
 % | |__   _  __ _ _   _ _ __ ___  ___ 
 % |  __| | |/ _` | | | | '__/ _ \/ __|
 % | |    | | (_| | |_| | | |  __/\__ \
 % |_|    |_|\__, |\__,_|_|  \___||___/
 %            __/ |                    
 %           |___/                     
 %

\ifx\modelvtwo\undefined
\begin{figure*}[ht]
\centering
  \includegraphics[width=\textwidth]{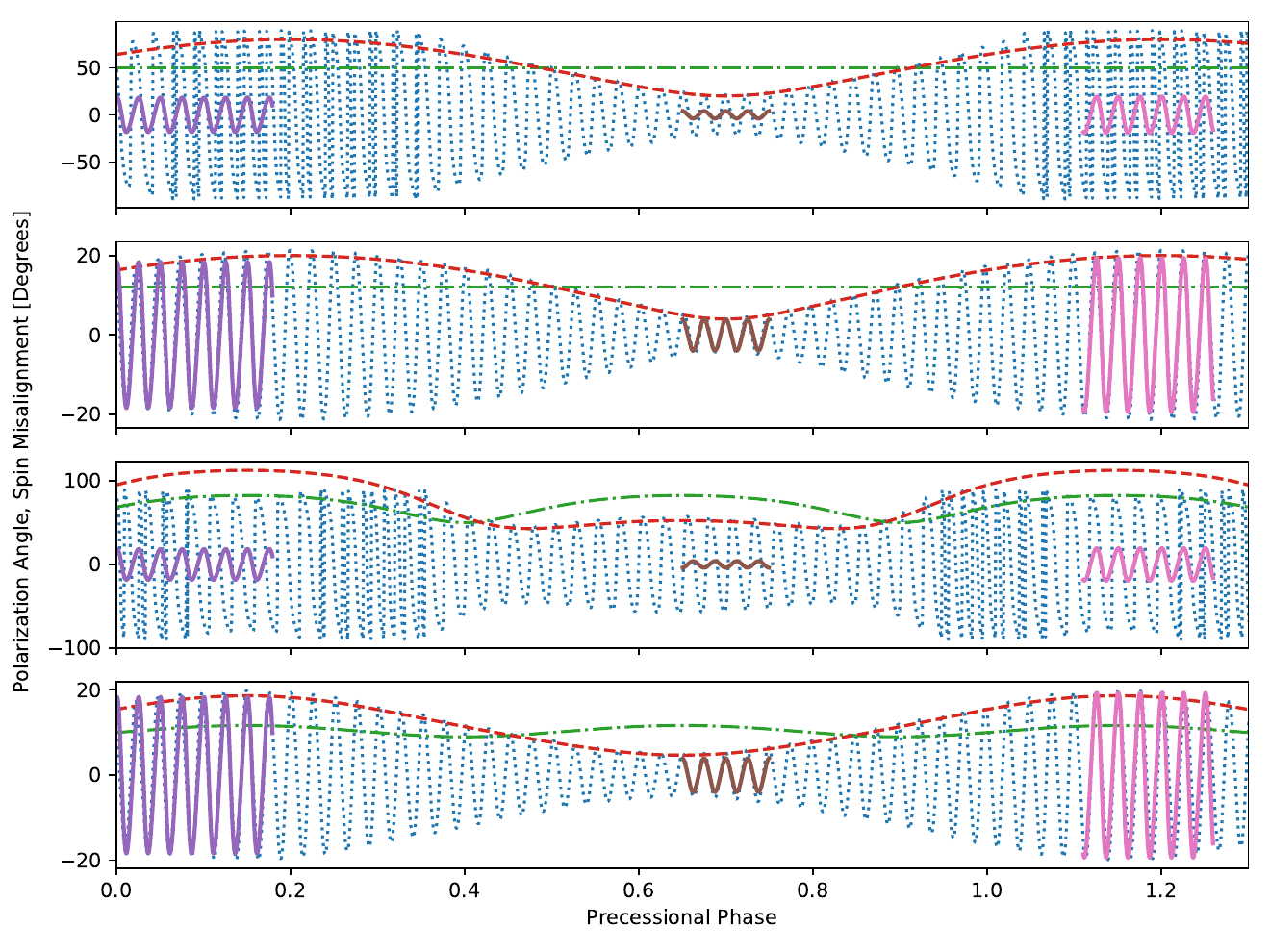}
    \caption{\textbf{Four models for the precession.} The models are depicted in blue dotted curves, and the best-fitting RVM PAs from the three epochs are superimposed in purple, brown and pink curves from left to right.  To show both the rotational evolution and the precession evolution as well as the duration of each observation over the precession, the rotational period has been increased from 1.2~s to about one day. Uppermost: a symmetric model\cite{2013MNRAS.435.1147P} with spin misalignment of $50^\circ$ and magnetic obliquity of $30^\circ$. Second: a symmetric model with spin misalignment of $11^\circ$ and magnetic obliquity of $8^\circ$. Third: a triaxial model\cite{2022MNRAS.513.3359K} with a magnetic obliquity of $30^\circ$. Lowermost: a triaxial model\cite{2022MNRAS.513.3359K} with a magnetic obliquity of $7^\circ$. 
    %The green curve traces the elevation angle of the inner edge of the accretion disc\cite{2000ApJ...539..392S}. 
    The green dot-dashed curve traces the spin misalignment angle $\alpha$, and the red dashed curve traces $\theta$, the angle between the spin axis and the magnetic axis.}
    %The orange curve traces the spin misalignment angle.}
    \label{fig:models}
\end{figure*}
\else
\begin{figure*}[ht]
\centering
  \includegraphics[width=\textwidth]{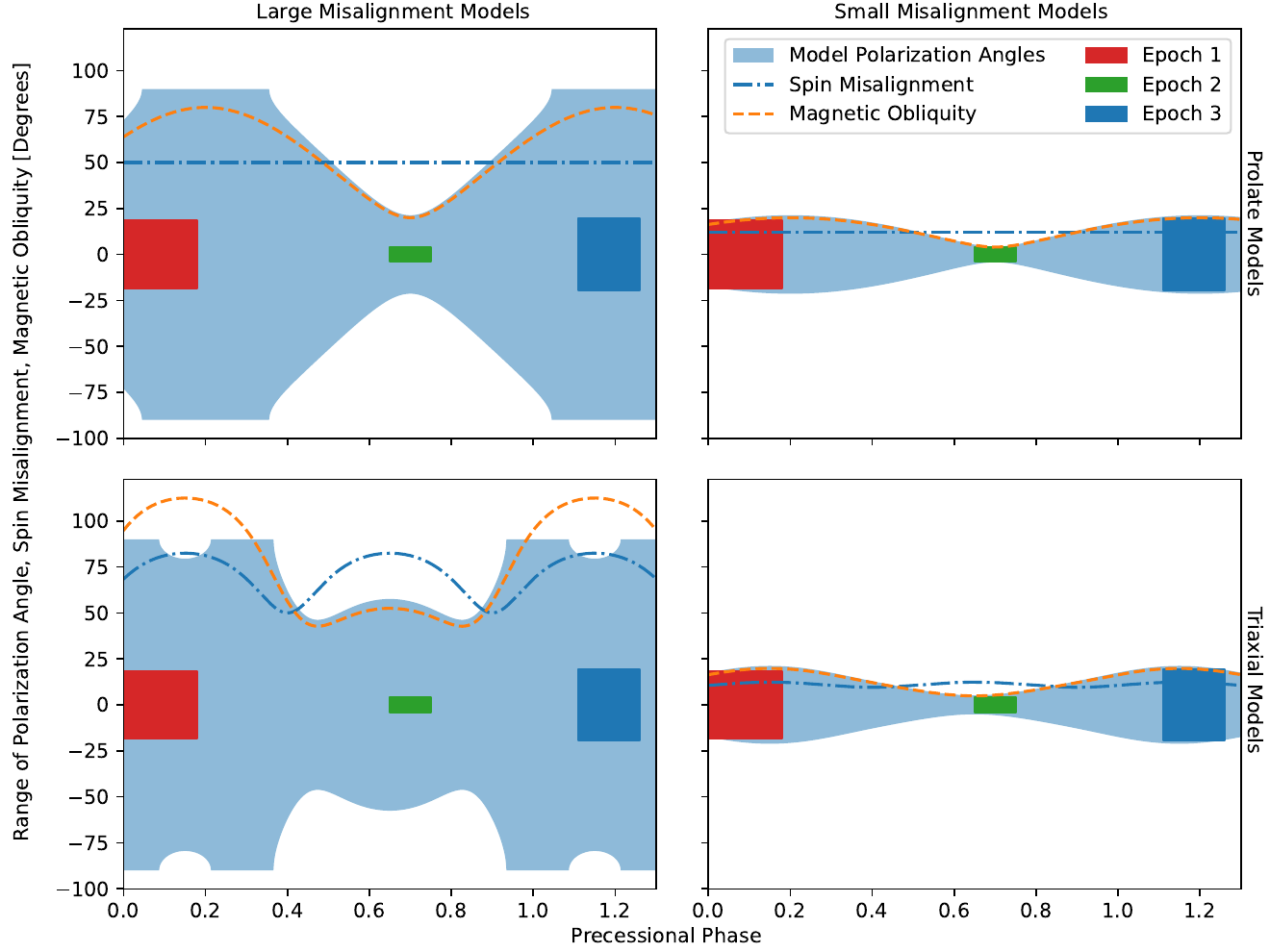}
    \caption{\textbf{Four models for the precession.} The extent of the polarization angle swing as a function of precessional phase for the four precession models is depicted by the light-blue bands, and the extent of best-fitting RVM PAs from the three epochs is superimposed in red, green and dark-blue bands from left to right.   Top Left: a prolate model\cite{2013MNRAS.435.1147P} with spin misalignment of $50^\circ$ and magnetic obliquity of $30^\circ$. Top Right: a prolate model with spin misalignment of $11^\circ$ and magnetic obliquity of $8^\circ$. Bottom Left: a triaxial model\cite{2022MNRAS.513.3359K} with a magnetic obliquity of $30^\circ$. Bottom Right: a triaxial model\cite{2022MNRAS.513.3359K} with a magnetic obliquity of $7.5^\circ$. 
    %The green curve traces the elevation angle of the inner edge of the accretion disc\cite{2000ApJ...539..392S}. 
    The blue dot-dashed curve traces the spin misalignment angle $\alpha$, and the orange dashed curve traces $\theta$, the angle between the spin axis and the magnetic axis.}
    %The orange curve traces the spin misalignment angle.}
    \label{fig:models}
\end{figure*}
\fi

% \begin{figure*}[ht]
% \centering
%   \includegraphics[width=\textwidth]{figures/deltanu.pdf}
%   \caption{\textbf{Spin Frequency Deviation.}  After the long-term trend of the spin frequency of Hercules X-1 is removed\cite{2022MNRAS.513.3359K} from the Fermi GBM frequency measurements, the variation with superorbital period becomes apparent.  The results from the second symmetric model in Fig.~\ref{fig:models} are superimposed in green, and those from lowermost model are shown in orange.  The  X-ray flux as observed by the Swift observatory is depicted by an olive curve along the bottom of the plot.}\label{fig:deltanu}
% \end{figure*}

\clearpage

 %  __  __      _   _               _     
 % |  \/  |    | | | |             | |    
 % | \  / | ___| |_| |__   ___   __| |___ 
 % | |\/| |/ _ \ __| '_ \ / _ \ / _` / __|
 % | |  | |  __/ |_| | | | (_) | (_| \__ \
 % |_|  |_|\___|\__|_| |_|\___/ \__,_|___/
     
\begin{methods}

To model the polarization direction as a function of rotational and precessional phase we use the formalism of Euler angles\cite{Landau1976Mechanics} to yield the following expression
\begin{eqnarray}
\tan(\textrm{PA}-\chi_M) &=& \Biggr [ {\sin{\theta_B } \sin{\varphi } \cos{\left(\psi - \psi_B \right)} + \sin{\theta_B } \sin{\left(\psi - \psi_B \right)} \cos{\varphi } \cos{\alpha } - }  \nonumber \\
& & { \sin{\alpha } \cos{\theta_B } \cos{\varphi }} \Biggr ] \Biggr [ 
{- \sin{i_M } \sin{\theta_B } \sin{\alpha } \sin{\left(\psi - \psi_B \right)} - } \\ 
& &  {- \sin{i_M } \cos{\theta_B } \cos{\alpha } - \sin{\theta_B } \sin{\varphi } \sin{\left(\psi - \psi_B \right)} \cos{i_M } \cos{\alpha } + } \nonumber \\
& & { \sin{\theta_B } \cos{i_M } \cos{\varphi } \cos{\left(\psi - \psi_B \right)} + \sin{\varphi } \sin{\alpha } \cos{i_M } \cos{\theta_B }} \Biggr ]^{-1} \nonumber , 
\end{eqnarray}
where $\chi_M$ is the position angle of the angular momentum on the plane of the sky, $\theta_B$ is the angle between the symmetry axis with the largest moment of inertia and the magnetic axis (the magnetic obliquity), $\varphi$ is the rotation angle, $\psi$ is the precession angle, $\psi_B$ is longitude of the magnetic axis using the symmetry axis with the smallest moment of inertia as a reference, $\alpha$ is the angle between the angular velocity and the symmetry axis with the largest moment, and $i_M$ is the angle between the angular momentum and the line of sight.  \ifx\deltanu\undefined
\relax
\else
The angles are depicted in Extended Data Figure~\ref{fig_extended:Euler}.
\fi

Because the asymmetry of the moment of inertia is small, the values of $\chi_M$ and $i_M$ are nearly constant in the absence of torques; furthermore, the evolution of the angles in the symmetric case where two of the moments of inertia are equal is straightforward
\begin{equation}
    \psi = 2 \pi \nu_\textrm{\scriptsize prec} t, \qquad
    \varphi = 2 \pi \nu_\textrm{\scriptsize rot} t, \qquad
    \alpha = \textrm{constant}.
\end{equation}
In the case where the star is oblate the two frequencies have opposite signs, and the precession is retrograde.  Even in the biaxial case, the observed rotation frequency defined as the inverse of the timescale between the crossings of PA through $\chi_M$, for example, is not constant through the precession (see Extended Data Figure~\ref{fig:deltanu})\cite{2022MNRAS.513.3359K}.  The value of $\nu_\textrm{\scriptsize rot}$ is the mean rotation frequency over the precession period.

The situation for a triaxial body is somewhat more complicated but also straightforward\cite{Landau1976Mechanics}.  Let us define 
the three eigenvalues of the moment of inertia tensor to be $I_3>I_2>I_1$.  If we let
$\alpha_{0}$ to the be minimum spin misalignment which occurs when the spin axis is nearest the axis with the smallest moment of inertia ($I_1$), we can define the following parameters
\begin{equation}
 k^2 = \frac{I_3 \left( I_{2}- I_{1} \right) }{I_1 \left( I_{3} - I_{2}\right) } 
 \tan^{2}\alpha_0 
 %\tan^{2}{\left(\alpha_0 \right)}
, \qquad n=   \frac {I_1 (I_3-I_2)}{I_2(I_3-I_1)} , 
\end{equation}
where the time parameter through the precession is given by
\begin{equation}
    \tau =  4 K(k) \nu_\textrm{\scriptsize prec} t  , 
\end{equation}
where $K(k)$ is a complete elliptic integral of the first kind (\texttt{ellipk} in \texttt{scipy}).  The precession rate is 
\begin{equation}
\nu_\textrm{\scriptsize prec}= \frac{\Omega_{3,0}}{4 K(k)}\sqrt{\frac{(I_3-I_2)(I_3-I_1)}{I_1 I_2}}     , 
\end{equation}
where $\Omega_{3,0}$ is the angular velocity about the minor axis at the time of minimum spin misalignment.

Using the Jacobi elliptic functions (\texttt{ellipj} in \texttt{scipy}) we have
\begin{equation}
    \tan\psi(\tau) = \sqrt{n}\ \frac{\textrm{cn}(\tau|k^2)}{\textrm{sn}(\tau|k^2)},~~ \cos\alpha(\tau) = \textrm{dn}(\tau|k^2)  \cos\alpha_0 , 
\end{equation}
where $\tan\psi$ is inverted to take account for the sign of the numerator and denominator so the range is zero to $2\pi$ (e.g.\ by using \texttt{arctan2}).  The Jacobi elliptic functions are periodic with a period of $K(k)$ so the functions $\psi(\tau)$ and $\alpha(\tau)$ are also periodic. The expression for the angle $\psi$ ensures that $\psi$ decreases in time, so the precession again is retrograde as expected. The final parameter is the rotational angle $\varphi$ which is given by\cite{Landau1976Mechanics}
\begin{equation}
\varphi=\left [2\pi \nu_\textrm{\scriptsize rot} t - \frac{2\pi\nu_\textrm{\scriptsize rot} }{\Omega_{3,0}} \frac{I_2-I_1}{\sqrt{(I_3-I_2)(I_3-I_1)}}\sqrt{\frac{I_2}{I_1}} \int_0^\tau \cos^2\psi\ d\tau'\right ] \left ( 1 - \frac{I_2-I_1}{I_1} \langle\cos^2\psi\rangle \right )^{-1}.
\end{equation}
and in the approximation where we neglect the differences in the moments of inertia beyond first order we have $\Omega_{3,0}\approx 2 \pi \nu_\textrm{\scriptsize rot}\cos\alpha_0$ and
\begin{equation}
    \varphi = 2\pi \nu_\textrm{\scriptsize rot} t - \frac{1}{\cos\alpha_0} \frac{I_2-I_1}{\sqrt{(I_3-I_2)(I_3-I_1)}}\int_0^\tau \left ( \cos^2\psi  - \langle\cos^2\psi\rangle \right ) d \tau' .
\end{equation}
These two integrals can be expressed in closed form using Jacobi theta functions with complex arguments\cite{Landau1976Mechanics}; however, it is straightforward to integrate them numerically.  Although the second term in this expression is explicitly periodic in $\tau$ with a period of $K(k)$, the first term is not, so the star does not necessarily return to the same configuration.

\end{methods}
\bibliography{herx1}

\begin{addendum}
    \item[Acknowledgements] 
The Imaging X-ray Polarimetry Explorer (IXPE) is a joint US and Italian mission.  The US contribution is supported by the National Aeronautics and Space Administration (NASA) and led and managed by its Marshall Space Flight Center (MSFC), with industry partner Ball Aerospace (contract NNM15AA18C).  
The Italian contribution is supported by the Italian Space Agency (Agenzia Spaziale Italiana, ASI) through contract ASI-OHBI-2017-12-I.0, agreements ASI-INAF-2017-12-H0 and ASI-INFN-2017.13-H0, and its Space Science Data Center (SSDC) with agreements ASI-INAF-2022-14-HH.0 and ASI-INFN 2021-43-HH.0, and by the Istituto Nazionale di Astrofisica (INAF) and the Istituto Nazionale di Fisica Nucleare (INFN) in Italy.
This research used data products provided by the IXPE Team (MSFC, SSDC, INAF, and INFN) and distributed with additional software tools by the High-Energy Astrophysics Science Archive Research Center (HEASARC), at NASA Goddard Space Flight Center (GSFC). JH acknowledges support from the Natural Sciences and Engineering Research Council of Canada (NSERC) through a Discovery Grant, the Canadian Space Agency through the co-investigator grant program, and computational resources and services provided by Compute Canada, Advanced Research Computing at the University of British Columbia, and the SciServer science platform (www.sciserver.org). 
D.G.-C.  acknowledges support from a CNES fellowship grant.  
JP and SST were  supported by the Academy of Finland grants 333112, 349144, 349373, and 349906 and the V\"ais\"al\"a Foundation. 
VD and VFS thank the German Academic Exchange Service (DAAD) travel grant 57525212.
%This work was supported by the Natural Sciences and Engineering Research Council of Canada. 
We used Astropy:\footnote{http://www.astropy.org} a community-developed core Python package and an ecosystem of tools and resources for astronomy.
%\begin{addendum}
    \item[Data Availability] 
    The data used for this analysis are available through HEASARC under IXPE Observation IDs: 01001899, 02003801 and 02004001.
    \item[Code availability] 
    The software used for this analysis is available at \\\texttt{https://github.com/UBC-Astrophysics/IXPE-Analysis}
%\end{addendum}
    \item[Author Contributions] 
    J.H. analysed the data and wrote the draft of the manuscript. J.P. led the
work of the IXPE Topical Working Group on Accreting Neutron Stars and
contributed to the interpretation and the text.
V.D., D.G.-C., I.C., A.M., S.S.T., D.M. and V.F.S. contributed to interpretation of the results and writing of the text. A.M. created Fig.~\ref{fig:schematic}. M.B. and G.G.P. acted as internal referees
of the paper and contributed to interpretation. Other members of the IXPE
collaboration contributed to the design of the mission and its science case and
planning of the observations. All authors provided input and comments on the
manuscript
    \item[Competing Interests] The authors declare that they have no competing financial interests.
    \item[Correspondence] Correspondence and requests for materials should be addressed to J.H. (email: heyl@phas.ubc.ca).
\end{addendum}

%
%   _____                   _                           _                   
%  / ____|                 | |                         | |                  
% | (___  _   _ _ __  _ __ | | ___ _ __ ___   ___ _ __ | |_ __ _ _ __ _   _ 
%  \___ \| | | | '_ \| '_ \| |/ _ \ '_ ` _ \ / _ \ '_ \| __/ _` | '__| | | |
%  ____) | |_| | |_) | |_) | |  __/ | | | | |  __/ | | | || (_| | |  | |_| |
% |_____/ \__,_| .__/| .__/|_|\___|_| |_| |_|\___|_| |_|\__\__,_|_|   \__, |
%              | |   | |                                               __/ |
%              |_|   |_|                                              |___/ 
%

\renewcommand{\figurename}{Extended Data Figure}
\setcounter{figure}{0} 

\begin{figure*}[ht]
    \centering
    \includegraphics[width=0.9\textwidth]{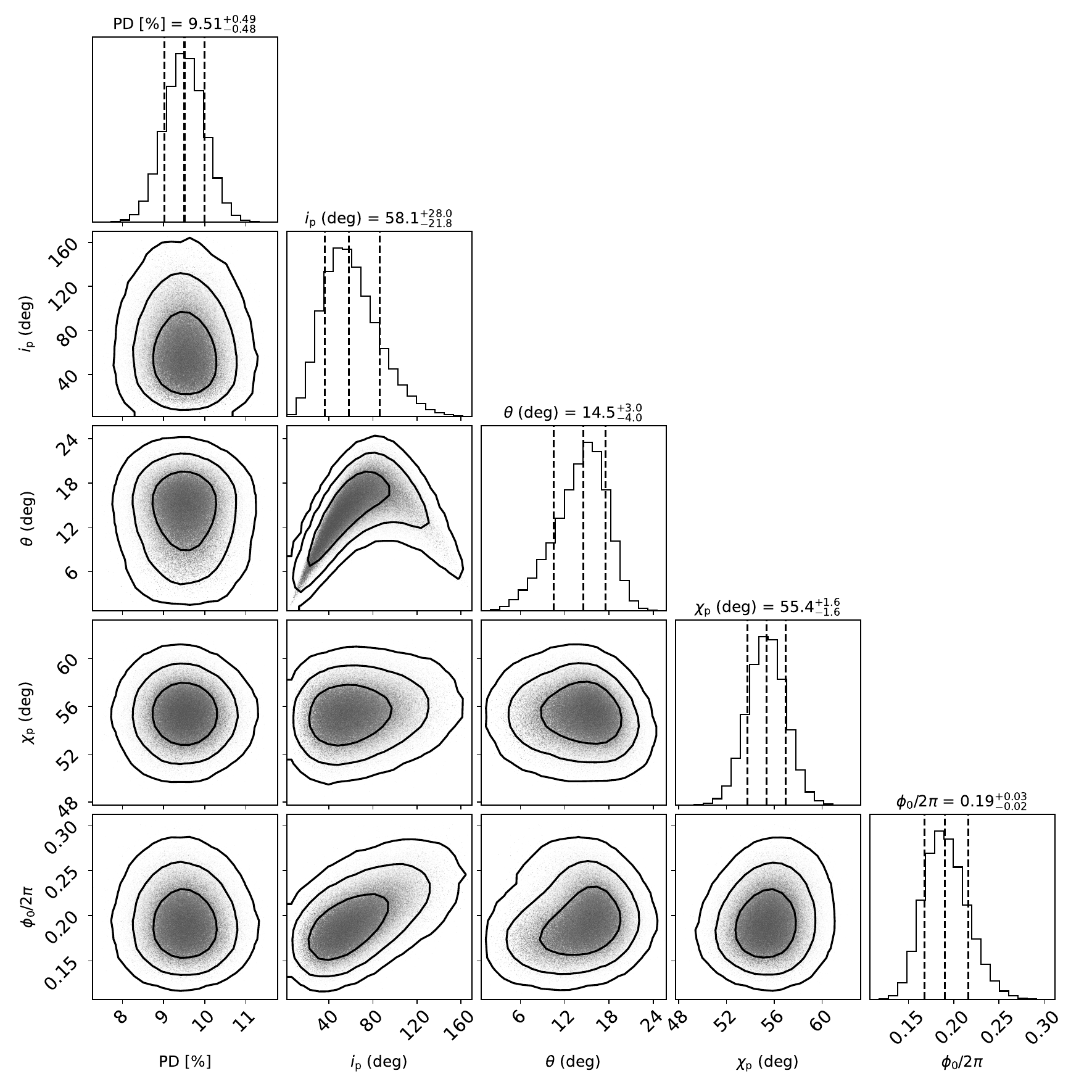}
    \caption{\textbf{Posteriors for the RVM for the First Main-On (2022 January)}  The two-dimensional contours correspond to $68\%$, $95\%$ and $99\%$ confidence levels. 
The histograms show the normalized one-dimensional distributions for a given parameter derived from the posterior samples.    } \label{fig_extended:epoch1}
\end{figure*}

\begin{figure*}[ht]
    \centering
    \includegraphics[width=0.9\textwidth]{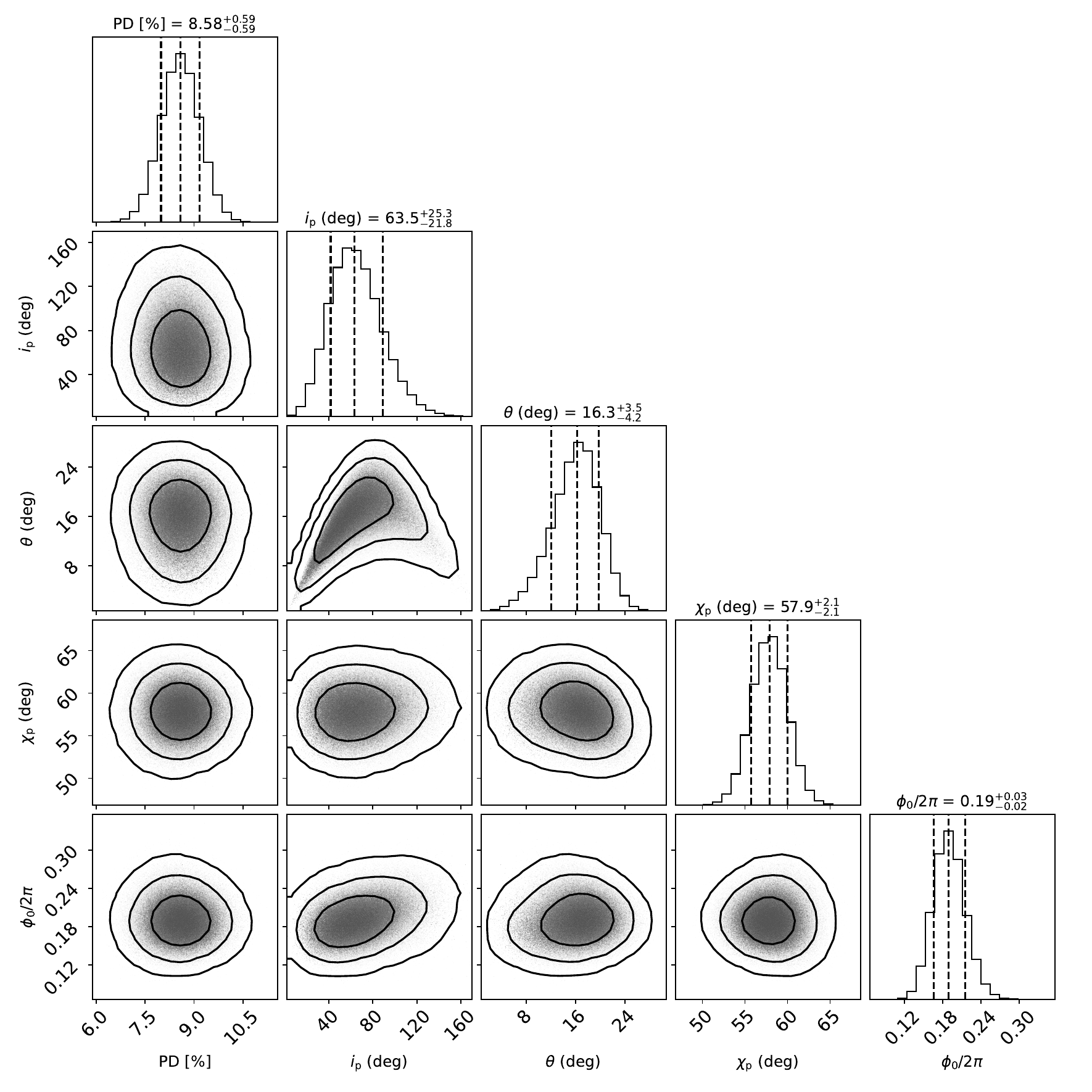}
    \caption{\textbf{Posteriors for the RVM for the First Main-On (early)}    } \label{fig_extended:epoch1-early}
\end{figure*}

\begin{figure*}[ht]
    \centering
    \includegraphics[width=0.9\textwidth]{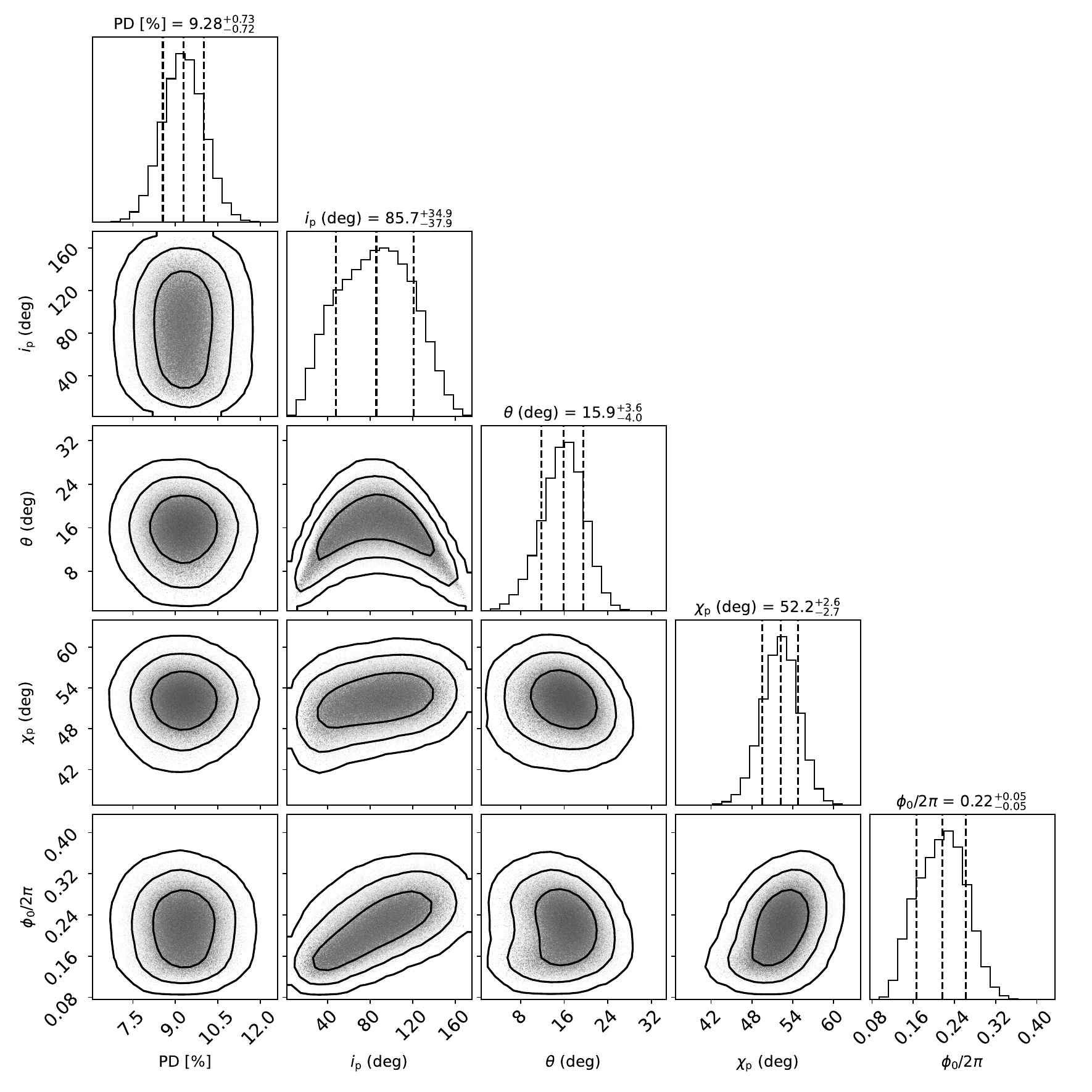}
    \caption{\textbf{Posteriors for the RVM for the First Main-On (late)}    } \label{fig_extended:epoch1-late}
\end{figure*}

\begin{figure*}[ht]
    \centering
    \includegraphics[width=0.9\textwidth]{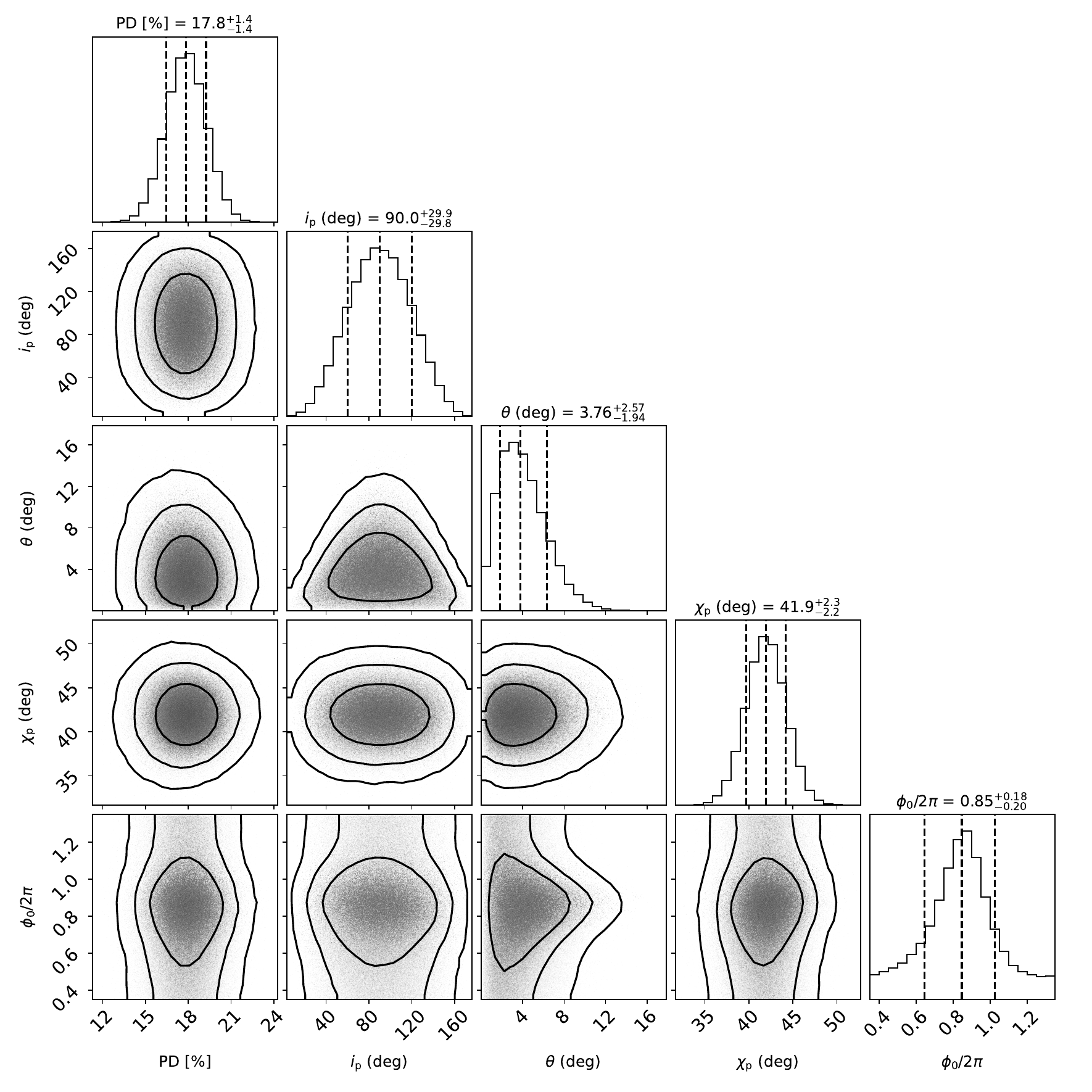}
    \caption{\textbf{Posteriors for the RVM for the Short-On (2023 January)}     }\label{fig_extended:epoch2}
\end{figure*}
\begin{figure*}[ht]
    \centering
    \includegraphics[width=0.9\textwidth]{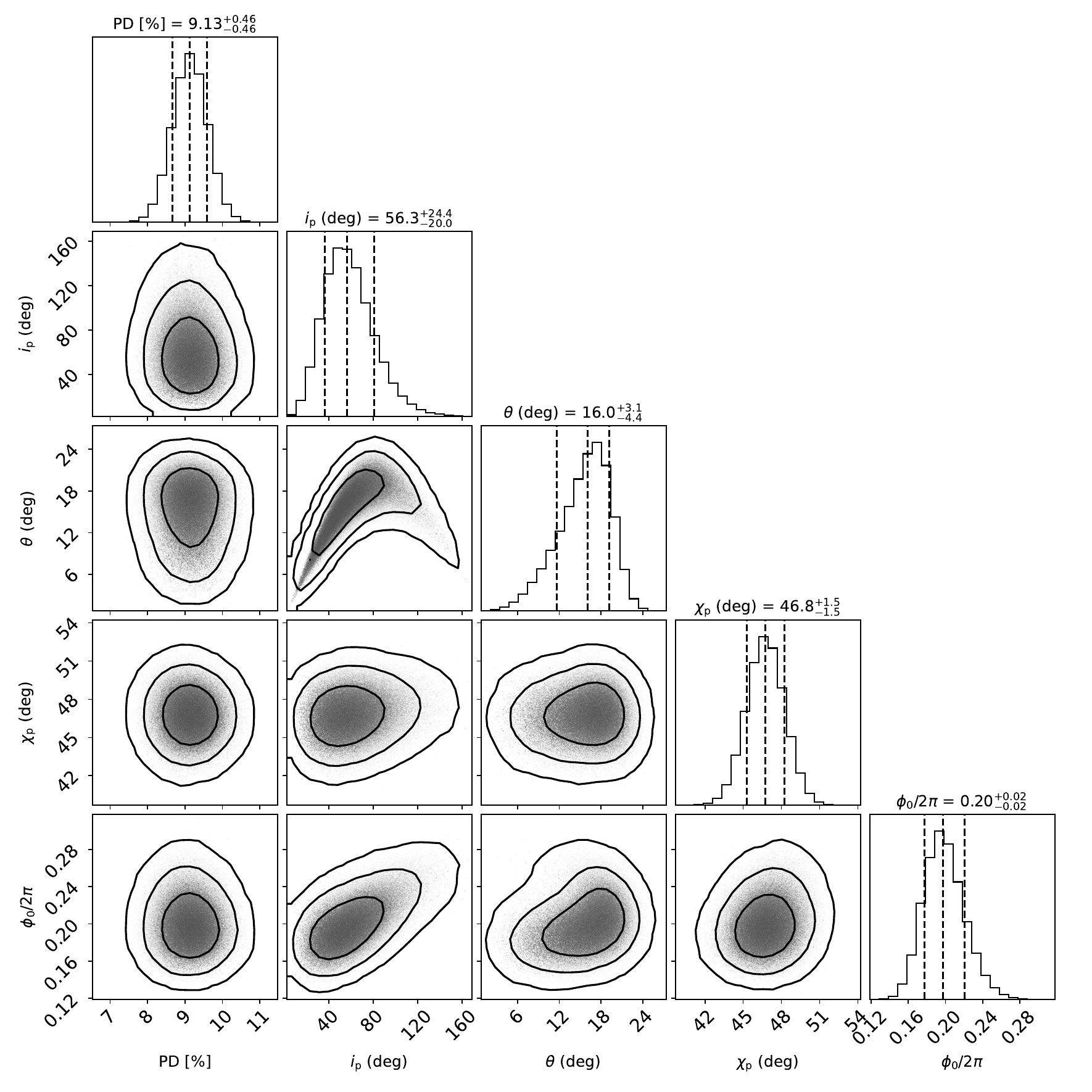}
    \caption{\textbf{Posteriors for the RVM for the Second Main-On (2023 February)}    } \label{fig_extended:epoch3}
\end{figure*}

\begin{figure*}[ht]
    \centering
    \includegraphics[width=0.8\textwidth]{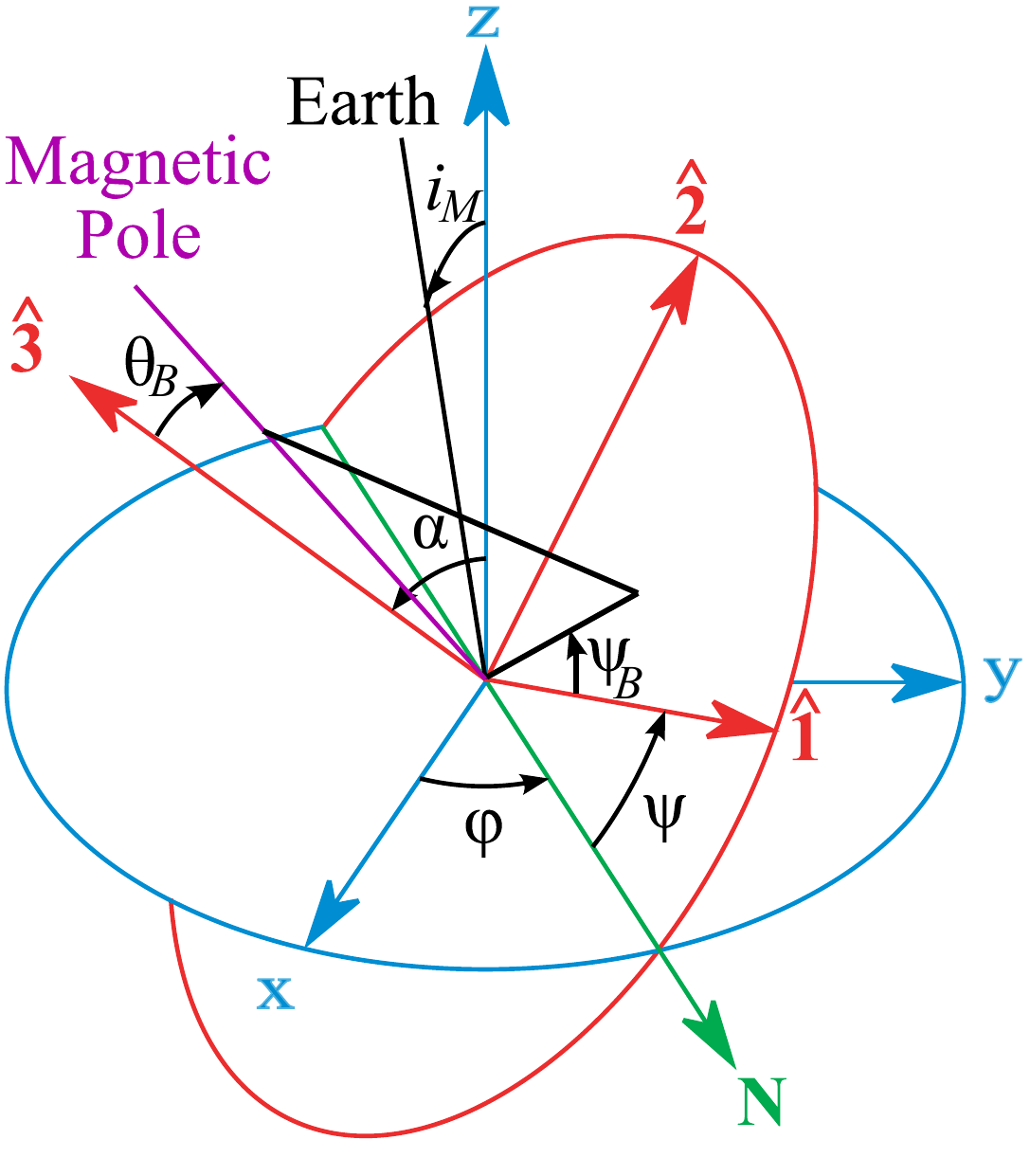}
    \caption{\textbf{Euler Angles as defined in the Methods.} The axes $\hat{1},\hat{2}$ and $\hat{3}$ denote the body axes of the neutron star crust, and the axes $x, y$ and $z$ denote a frame of reference with the direction to Earth fixed. The line of sight is indicated by the black line labelled ``Earth'', makes an angle $i_M$ with the $z$-axis and lies in the $x-z-$plane.  The magnetic pole is indicated by the purple line, makes an angle $\theta_B$ with the $\hat 3-$axis and lies in the $\hat{2}-\hat{3}-$plane. The vector labeled N denotes the line of nodes about which the rotation by $\alpha$ is performed.} \label{fig_extended:Euler}
\end{figure*}

\ifx\deltanu\undefined
\relax
\else
 \begin{figure*}[ht]
 \centering
   \includegraphics[width=\textwidth]{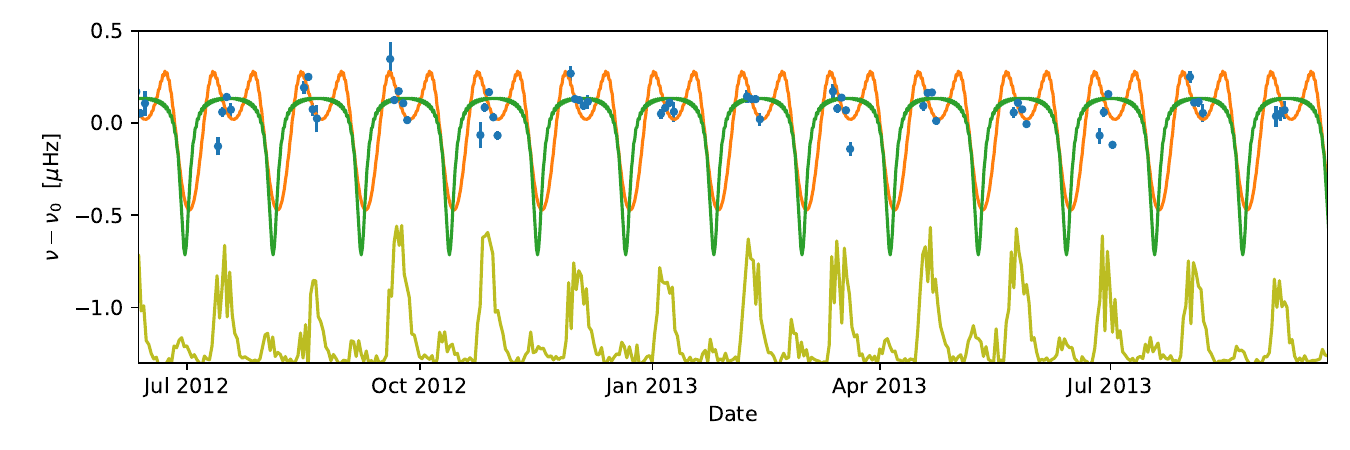}
   \caption{\textbf{Spin Frequency Deviation.}  After the long-term trend of the spin frequency of Hercules X-1 is removed\cite{2022MNRAS.513.3359K} from the Fermi GBM frequency measurements, the variation with superorbital period becomes apparent.   The curves are not a fit to the data, but rather the expectations from the two small-misalignment precession models outlined in the text. The results from the prolate model in Fig.~\ref{fig:models} are superimposed in green (single hump), and those from triaxial model are shown in orange (double hump).  The xX-ray flux as observed by the Swift observatory is depicted by an olive curve along the bottom of the plot.}\label{fig:deltanu}
\end{figure*}
\fi

\end{document}